\documentstyle[aps,psfig,epsfig]{revtex}

\renewcommand{\=}{~=~}

\begin{document}
\draft
\twocolumn

\title{Realistic modellization for the growth of a yeast colony.}
\author{Lorenzo Fortunato}
\address{Dipartimento di Fisica "G.Galilei" and Istituto Nazionale di Fisica 
Nucleare di Padova, \\ v. Marzolo 8, 35131 Padova, Italy }

\maketitle

\begin{abstract}
An evolving yeast colony is simulated by means of a cellular automaton that takes
 care of many important features of the system under study. A complete survey of 
the properties of the colony is done and a set of scaling relations is found, 
together with the analysis of the critical exponents. The Family-Vicsek
relation between them is verified to a good accuracy.\\ 
A mobility parameter is
introduced in order to relate the behaviour of the growth exponent with the 
temperature and we propose that the brownian motion is a key feature to describe
the bacterial and fungi growth process in diluted solutions.\\
The chemical and geometrical correlations are found to play a relevant role in 
the modellization, so that the need of realism is emphasized. A new phase space 
built upon the relevant 'thermodinamical' variables is explored.
\end{abstract}

In the realm of biology the importance of growth of bacterial and fungi colonies 
\cite{fun},\cite{Mats},\cite{Ben},\cite{Sams},
as well as the tissue formation in hystology \cite{hys} or  the evolution of species
\cite{evo}, are well established field of development.
 When dealing with epidemiology \cite{epi} or front propagation \cite{for} 
 computer simulations inspired to "lattice animals" or Cellular Automata \cite{CA} are 
 widespread implementations  for these very different systems.\\
Furthermore the origins of life are understood on an evolutive basis, starting from simple 
molecules arranging themselves in more complicated ones.  A primitive "first"  bacterium 
in a small puddle, in which nutrient substances are dissolved or  a brewer that is seeding 
some yeast in his malt liquor (the so-called 'wort' ) to produce beer  are very different
examples of the systems we are intended to study.\\
We think also to the technological importance of yeast in the brewing or baking process, 
or more in general in  fermentation processes.\\
The Eden model \cite{eden} is often cited when the growth process
is discussed: actually the shape of the clusters remind the observed ones, but the 
scaling behaviour is far from the measured one and hence things are much more involved.  
Furthermore to show the capability of computer simulation to reach a stable state is cited
the Convey's Game of Life  \cite{GoL}, that could have something in common with the 
evolution of a bacterial colony,  but it is far from be realistic.
These and others early models were often based on simple duplication laws, 
concerned, for example, only with the number of neighbouring unoccupied sites.\\
From all these considerations we are faced with the problem of  interpreting correctly and
modelling simple biological or ecological systems. The main issue of this work is to 
emphasize the fact that is important to include all the relevant details to generate a 
complex enough behaviour. 
It has been suggested \cite{Ben} that chemical correlations are necessary to explain the 
very different morphologies and evolutive features of growing living systems.
We introduce in our model 'instantaneous' chemical correlations
represented by the concentration fields of nutrient and waste and we explore systematically a 
new phase diagram that relates the number of generation of a colony to the above cited 
concentrations. The mobility of the cells, that is a function of the temperature,
is shown to play a relevant role in the evolution of such a system. We measure power 
laws with the purpose to study their modifications when the yeast is 
allowed to wander randomly.
This work starts from that foundations with the intent to model properly the system under
investigations (Sec. II), to generate accurate patterns and to measure 
some important quantity (Sec. II-III) giving estimates of scaling exponents. Conclusions 
on the results are drawn in Sec. IV.

\section{The system  and the model.}
The complex system we are studying needs to be implemented taking care of all the
 necessary details. We will consider a Cellular Automaton  (CA) defined on  a square lattice
(due to the limitations of memory the maximum side length is about
 800-1000 units)  with periodic boundary conditions. The simulations
start all from a single seed placed in the center of the grid. The behaviour of
every cell depends on the eight nearest neighbours: this is the Moore type
neighbourhood as defined in \cite{Wolf}.\\
The typical yeast life cycle is very simple to mimic and we examine the simplest
behaviour that a cell could keep: it eats nutrient (a generic name that means 
above all sugar, with small percentage of other substances) and it eliminates waste 
(that in our context is alcohol).
The specific chemical composition of these substances is however not important here. \\
We define two overall  initial variables: the concentration of nutrient $C_{food}^{init}$
and the concentration of waste $C_{waste}^{init}$ (usually set to zero, but one could 
wish to explore other regions in the  $C_{food} - C_{waste}$ phase space).
These quantities are not
properly concentrations but rather amount of solute because they are defined as the 
product of the squared size of the grid (the volume) and the percentage (a number 
from 0 to 1). \\
The reason to introduce such variables is simple: in the real 
situation yeast cells cannot tolerate an alcohol concentration higher than $12\%$.
The colony is "killed" by poisoning when this value is reached: this is some sort of
criticality because  the system grows, obeying to certain probability rules that
we will explain below, consuming food and producing waste until the refusals
are too much to let the colony live. For our purposes the initial concentration of
food is very large acting as a reservoir, but we can imagine situations in
which the colony is "killed" (or properly inactivated) by starvation.
One may think to other strategies to implement food and waste like introducing a
food cell that is absorbed when in contact with a yeast cell, thus producing one 
waste cell. This is however not necessary because the typical duplication time of a
colony is of the order of 30-60 minutes, depending on various strains of yeast, on the
temperature and other environmental features.
In that amount of time, at room temperature and
for small enough systems, the solution has the same concentration in every point
because a normal yeast cell is an oval of about $5\mu m$ and hence,
accordingly with what we have stated before, the maximum grid could have a size of
about $800 \cdot 5 \mu m \= 4 mm$, some millimeters in the biggest case.  \\
 The larger is the area exposed to the solution, the higher is the exchange rate
 both of nutrients and of waste. Hence if all the nearest neighbouring cells are 
 occupied there is no way to eat: in such condition the yeast survives but is in 
 a suspended situation. If we allow  the CA to move, as explained below, the
 yeast cell can restart to have a normal life in any moment in the future. This takes care 
 of the effect of overcrowding in the evolution of the system.\\
We can imagine that this movement is characterized by a simple random walk of
one step in one of the eight surrounding squares (like the King movement in the
game of chess). A "mobility" parameter $T_p$ is related to the average brownian
motion of the full points in the grid. For every generation a cell is chosen at
random and moved randomly in one of the eight directions (if it is not occupied by
another member of the community). Obviously a completely surrounded one cannot
move, but it could return to walk in any successive step of the colony evolution.
  This is repeated $[N T_p]$ times where $N$ is number of individuals and $T_p$
is the mobility parameter, that becomes an integer when we take the floor function as 
indicated with square brackets. In so doing a given cell could be picked out more 
than one time, and there could be cells that are not moved. This is intended to 
portray a completely random environment and is in some way a new type of modellization
 for such a specific system. Similar stochastic automata are discussed with very
 great care for example in Ref.\cite{sincro}, \cite{update}. \\
An important point here is the introduction in our CA of a probability rule for
replication that depends upon all the relevant geometric and chemical details
included in the present discussion. At a certain generation $k$, that corresponds 
to a certain time $t_k$, the probability for an empty cell $\vec r$ of becoming occupied,
because of the mitosis of a neighbouring occupied cell, is 
\begin{equation}\label{prob}
p(\vec r,t_k)\= \underbrace{n_{emp} \over 8}_A ~\cdot \underbrace{C_{food}(t_k) 
\over C_{food}^{init}}_F 
~\cdot \underbrace{(C_{waste}^{death}-C_{waste}(t_k))
 \over C_{waste}^{death}}_W ,
\end{equation}
where $A$ is an area factor that is proportional to the number of empty neighbouring 
cells $n_{emp}$,
$F$ is a 'food' factor proportional to the overall concentration of food,
normalized to the initial value. It is 
one at the beginning and decreases during the evolution. $W$ is a 'waste' factor that
is zero at the first cycle and increases proportionally to the quantity of alcohol expelled.
The functional dependence of these factors on the concentrations has been taken to 
be linear, but one can think to other ways of implementing $F$ and $W$.
From the definitions above we see that $0\le p\le 1$.\\  
We denote by $c$  the amount of food absorbed by a single cell in every duplication
time, that is equal, for the sake of simplicity, to the amount of waste produced at the
same time. Hence we can write the concentrations of food and waste at the time $t_k$
as follows:
\begin{equation}
C_{food} (t_k) \=C_{food}^{init} ~-~ \biggl{(} \sum_{i=1}^{k-1}N_i \biggr{)}~c~,
\end{equation}
\begin{equation}
C_{waste} (t_k) \=\biggl{(}\sum_{i=1}^{k-1}N_i \biggr{)}~c~.
\end{equation}
$N_i$ is the population at the i-th cycle.\\
The whole life cycle is iterated thus generating the complex dynamics. A typical run 
consists of:
\begin{enumerate}
\item{Seeding of the 'first' progenitor.}
\item{Iteration of many generations as long as saturation is reached.}
\item{Measure of saturation values for radius and width.} 
\end{enumerate}
In particular for the point 2: every generation means a complete life cycle for the colony. 
The principal rules are the following:
\begin{itemize}
\item{Metabolism: every yeast cell eats an amount $c$ of food thus producing an equal
 quantity of waste.}
\item{Mobility: if $T \ne 0$ then every cell is moved accordingly to the stochastic rule
 exposed in the above discussion.}
\item{Duplication: if the probability $p(\vec r,t_k)$ is greater than a random
 number, chosen in the range $0-1$, a cellular subdivision occurs (mitosis).}
 \item{Measurement: the number of individuals and concentrations as well as
 radius and width are recorded.}
\end{itemize}
Many runs are typically taken into account, and extensive simulations are performed
with different sets of parameters, as will be discussed below.

\section{Discussion.}
Our simple automata are suitable to breed a huge variety of situation, but the gross
feature of the evolution in essentially the same: after the well-known initial 
exponentially growing phase the duplication becomes hindered by the cooperation of 
diverse mechanism, each leading to a small rate of birth of new individuals: 
overcrowding, starvation and poisoning act in the same direction. Note that this concepts
are not only effective in yeast growing, but a huge class of ecological and biological
systems is based on the same foundations, first of all the human kind, even if the
control mechanisms are much more intricately involved and hard to analyze.
The last part of the evolution reaches a saturation value after which the growth stops
completely because the alcohol concentration turns to the critical value. 
\subsection{$T_p =0$ case, no thermal motion.}
Let us begin with the discussion of the case with no brownian motion $Tp=0$, when cells
are not allowed to move on the grid. The shape is very compact and the colony develop
in a larger number of generation, until the poisoning is achieved. A typical evolution
leads to the behaviour showed in figure 1.
\begin{figure}
\begin{center}
\epsfig{figure=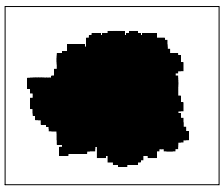,scale=.7}
\end{center}
\label{colonia0}
\caption{Shape of a colony at the saturation. The grid size is 200 and the temperature 
parameter is set to zero. The resulting shape is very compact and no cell could exist 
without a neighbour (or better without a parent).}
\end{figure}  
We have measured the number of cellular automata at 
every cycle, as well as the concentration of food and waste. 
The radius of gyration of the cluster is defined as:
\begin{equation}
R\= {\sum_{x,y} d(x,y) \over N_S} ~,
\end{equation}
where $d$ is the distance from the center of mass $(x_b,y_b)$ and  $N_S$ is the number 
of individuals in the surface region.
The semi-width of the interface between the colony and the ambient solution is:
\begin{equation}
w\= \sqrt{\sum_{x,y}( d(x,y)-R)^2 \over N_S} ~,
\end{equation}
These formulas are written in analogy with the correspondent quantities in a ballistic 
deposition simulation (see, for example, \cite{bara}). \\ In figure 2 
we report the width divided by $L^\alpha$ as a function of the time divided by $L^z$. 
The collapse is so good that one can hardly distinguish the four different symbols 
used for four different values of the grid size.
\begin{figure}[!t]
\label{collapse}
\epsfig{figure=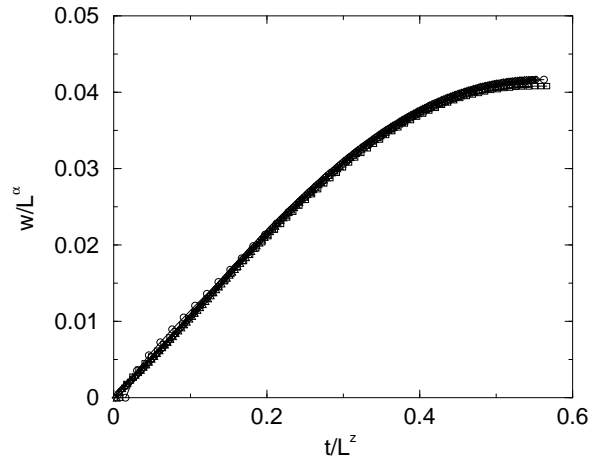,scale=0.6}
\caption{Data collapse for different values of the grid size: 100,200,500,800 (circles,
squares, diamonds and triangles respectively).  }
\end{figure}
The exponents $\alpha$,
$\beta$ and $z$ are defined measuring the scaling relations:
\begin{equation}
w_{sat} \sim L^\alpha  \quad ,\quad w_{sat} \sim t^\beta
\end{equation}
\begin{equation}
t \sim L^z , 
\end{equation}
with critical exponents \\
\begin{center}
\begin{tabular}{ccc} 
\hline 
~~&$T_p=0$\\
\hline
$~\alpha~$ &~0.961 ($\pm$ 0.005)~ \\
$~\beta~$ &~1.055 ($\pm$ 0.018) ~\\
$~z~$&~0.909 ($\pm $ 0.017)~\\
\hline
\end{tabular}
\end{center}
The errors considered here are only due to the linear fit. The relative importance
 of the statistical error measurements in the code could be estimated to be of the 
 order of $10^{-3}$ and it has been neglected. They are not shown in the loglog plots
  in figure above. Incidentally, in the $T=0$ case, we note that the 
Family-Vicsek scaling relation (see for reference \cite{bara})  holds almost perfectly:
$$ z\= {\alpha \over \beta} \= { 0.961 \over 1.055} \= 0.911 (\pm 0.016)~.$$ 
With the fitted values for the exponents we have checked the data collapse: in the next figure 
we report the width divided by $L^\alpha$ as a function of the time divided by $L^z$. 
The collapse is so good that one can hardly distinguish the four different symbols in the next
figure used for four different values of the grid size.

Our value are in good agreement with the few  existing issues in literature \cite{expe,bara}.
Vicsek gives an estimate for the exponent $\alpha_{exp} \= 0.78$ that is different 
from ours. However it must be said that the geometry of the experiment is rather 
different: they check the evolution of a colony on an agar-agar solid substrate 
inoculating the bacteria with an infected rod. In this way they don't have thermal 
motion of bacteria in appreciable way (a feature that we will address below) and the 
chemical correlations are less pronounced being the 'almost instantaneous' dilution 
of substances prohibited in such an arrangement. Furthermore we want to mention 
another possibility, not previously taken in to account: the system is not really 
two-dimensional. Some cell is budded above the colony thus forming a 2 or 3 layers system. 
This, in our opinion, could led to a reduction of some critical exponent.

 \subsection{$T_p \ne 0$ case, thermal motion.}
Now we turn to the description of the movements on the grid. In Fig. 3 we show that the 
brownian motion affects the width of the surface, as well as the size of the colony.
We expect that an 'hot' colony feels much more the finite size effects, and its scaling exponents
might change with the mobility parameter.
\begin{figure}
\begin{center}
\epsfig{figure=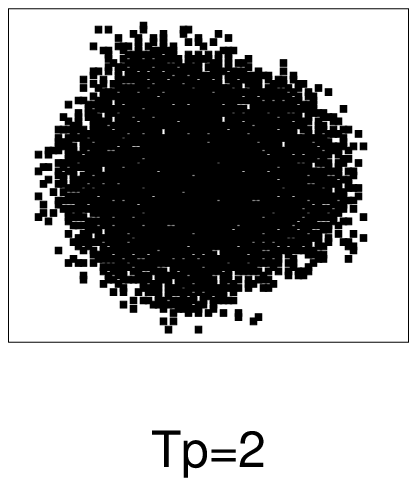,scale=.7}
\end{center}
\label{colonia2}
\caption{Shape of a colony at the saturation when the grid size is 200 and the 
mobility parameter is 2. The resulting shape has a diffuse width and there are holes 
in the interior. This is the effect of the brownian motion.}
\end{figure}
Obviously the surface width is greater than in the case of zero temperature and this is due
to the introduction of brownian motion. Every cell, and in particular a newly budded one,
could now move, thus allowing the surrounding cells to be in contact with the environment. 
This implies that they may eat and reproduce again.\\
Let us open a parenthesis on the role of $T_p$, before the analysis of the critical exponents.
To relate our mobility parameter to the real temperature we perform a number of simulations 
not allowing the cell to reproduce. The Einstein's relation \cite{Ein} for the diffusion 
could be used to 
relate the square of the diffusion length $<x^2>$ covered by the solitary cell with the 
interval of time $\Delta t$ that is a multiple of the time of a life-cycle:
\begin{equation}
<x^2> \= 2 D \Delta t ~,
\end{equation}
where $D$ is the diffusion coefficient expressed in terms of the temperature $T$, the
viscosity coefficient of water $v\=0.001~ (SI ~units)$, $N_{Av}$ is the Avogadro number,
$R$ the gas constant and the radius of the particle $r$:
\begin{equation}
D \={R T \over 6 \pi N_{Av} v r } ~.
\end{equation}
Refer again to \cite{Ein}, taking care of inserting the Stokes formula in the denominator.
Hence one obtains:
\begin{equation}
T \sim {<x^2> \over \Delta t} {3 \pi N_{Av} v \over R} r 
\sim{<(\Delta L)^2> \over n_g} {3 \pi N_{Av} v \over R t_g} r^3 ~,
\end{equation}
where the second expression is rewritten knowing that the distance over the grid 
$<(\Delta L)^2> \= <x^2> \slash r^2$ is measured taking the size of the grid $r$ 
as a unit, and that the time elapsed during the simulation is measured taking the period
of a generation $t_g$ as a unit, multiplied by the number of generations $n_g$.
 This is a quite coarse procedure.\\ 
Measuring the quantities for the first ratio and giving an estimate of the radius of
a bacterial cell are the only hurdles to overcome. While the code permits to achieve 
easily this ratio,  unfortunately the size of a yeast cell could vary between 2 
$\mu m$ and 10 $\mu m$, thus making the expectations for $T$ rather rough. 
It could be said, however, that a $T_p$ of a few units could led to a reasonable 
temperature.\\
The critical exponents are measured in a couple of case:
\begin{center}
\begin{tabular}{ccc} 
\hline 
~~&Tp=1&Tp=2\\
\hline
$~\alpha~$ &~0.931 ($\pm$ 0.012)~&~ 0.910 ($\pm$ 0.010)~\\
$~\beta~$ & ~1.096 ($\pm$ 0.010) ~&~ 1.113 ($\pm$ 0.004)~\\
$~z~$&~ 0.850 ($\pm $ 0.016)~&~ 0.818 ($\pm$ 0.010) ~\\
\hline
\end{tabular}
\end{center}
A general trend is the reduction of the $\alpha$ exponent with the temperature, 
at the same time $\beta$ increases and $z$ decreases.
Even if 
our geometrical set up is very different from the already cited experiment of 
Vicsek and even if we are dealing with diluted solutions rather than agar-agar 
plates, the observed reduction indicates that this is an essential feature to 
reproduce the reported experimental values. The only problem to solve is the delicated 
relation between the mobility of a typical yeast cell and our mobility parameter. 
We have made attempts in this direction and it seems that a $T_p~=~8~10$ generates a good
accordance with the cited experimental value.\\
We suggest that the value $0.78$, that is somewhere in the middle between the
Eden model and ours, could be obtained if one accepts the role of thermal 
brownian motion. The data for $\alpha$ systematically tend  towards lower 
values with increasing mobility parameter.\\
Even in these cases the Family-Vicsek relation is
verified, being the calculated $z$ very similar to the measured one:
$$T=1 \qquad ~~~{\alpha \over \beta} \= 0.849 (\pm 0.013) $$
$$T=2 \qquad ~~~{\alpha \over \beta} \= 0.818  (\pm 0.009) $$
The Family-Vicsek scaling relations seem to be verified to a good accuracy and even in 
these cases a data collapse similar to the previous one has been verified.\\

\section{Phase-space wandering.}
The chemical correlations established by the diffusion of substances in the
solution is studied by changing the initial parameters.
Let us play with a phase space in which the relevant axis are $N,
C_{food}$ and $C_{waste}$, fixing the mobility parameter to zero for semplicity.
Two regions can be identified accordingly to 
whether the colony is able to live or it must die. The localization of the
 surface between the 'life-zone' and the 'death-zone' is interesting. \\ 
The following figure illustrates the dependence of the number of individuals 
with the amount of food at disposal varied between $0-1$ for two different system size.
 As one can imagine $N$ growths if the initial amount of food is high, but the 
mathematical formula could not be simple to guess. We believe that it growths like a 
logarithm.
\begin{figure}[!b]
\epsfig{figure=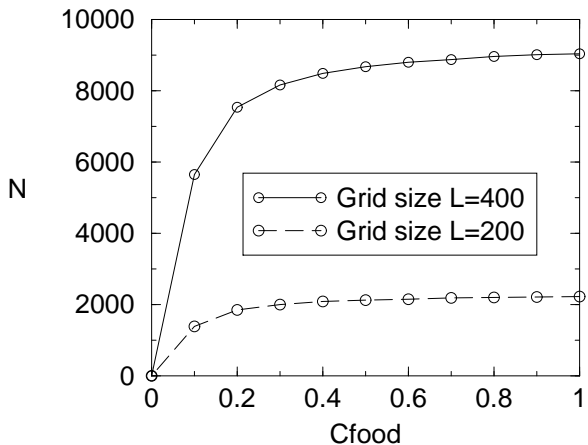,scale=0.7}
\caption{$\{ N - C_{food} \} $ Section of the phase space. Number of individuals vs. initial amount 
of food. The data are displayed for two different grid sizes.}
\end{figure}
The same has been done by keeping the amount of food fixed (and equal to $1$) and
changing the critical value for the waste from $0.12$ to $0.96$, for two different
grid sizes. The result are very similar, but the slope is less steep. However, 
as one could expect, the number of individuals is much more sensitive on this
parameters because this is the mechanism that actively stops the growth.
\begin{figure}
\epsfig{figure=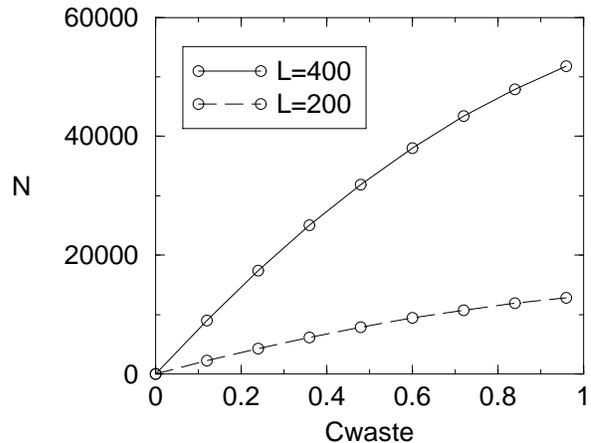,scale=0.7}
\caption{$\{ N - C_{waste}\} $Section of the phase space. Number of individuals vs. critical amount 
of waste. The data are displayed for two different grid sizes as indicated in
the figure.}
\end{figure}
Having this two sections in mind, one can easily reconstruct a surface in the three dimensional 
space $\{ N,C_{food},C_{waste} \}$, even if it's not simple to find the other section, 
namely the one in the  $\{ C_{food},C_{waste} \}$ plane,  
because the code does not allow to fix the number of cells that is an output.\\
Another interesting graph we want to discuss displays the behaviour of 
the number of generation $n_g$ with the amount of food. Surprisingly when there is a
small quantity of nutrient in the solution, the number of generation is greater.
\begin{figure}
\epsfig{figure=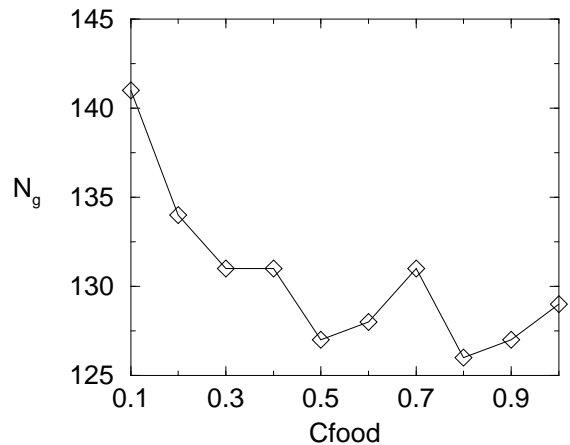,scale=0.7}
\caption{Number of generation that a colony develops before being killed by poisoning.
The x axis is the initial amount of food. The grid size here is 400 and the killing 
amount of alcohol is 12\%, as usual.}
\end{figure}
This is due to the fact that a few individuals are allowed to live, but not to
reproduce thus slightly consuming the food storage and producing only a few poison.\\
For intermediate values of food amount there are oscillations in the number of
generation, probably due to the effect of balance between the two regimes.\\
A similar, but much less impressive plot could be done by changing the killing amount of
waste and recording the number of generations.
\begin{figure}
\epsfig{figure=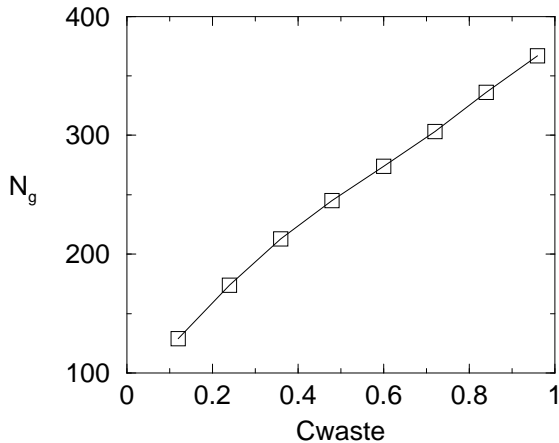,scale=0.7}
\caption{Number of generation that a colony develops before being killed by 
poisoning. The x axis is actually the amount of alcohol that stops the growth.
The grid size here is 400 and the initial amount of sugar is 100\%, as usual.}
\end{figure}
In this case the curve is almost linear and very easy to discuss: if a yeast cell could 
bear a larger concentration of poison in the solution, the number of generations growths
linearly. Different yeast strains have different tolerance with respect to alcohol
concentration and our model might be too simple to be of help when very different species are
considered. We think that the behaviour that we have discussed depends, to a large extent,
on the functional form of Eq. (1.1).

\section{Conclusions.}
We have depicted a typical scenario that holds for a large number of complex systems 
involving living organism. 
An accurate modelling has been achieved by means of a cellular automaton defined on 
the square lattice. The generated patterns as well as some important critical exponents 
have been achieved. The role of the brownian motion has been discussed and has been 
outlined the importance of a much deeper understanding of the behaviour of the exponents with
respect to the temperature that seems to influence the dynamic in a 
cumbersome way.\\
It is very probable that a proper understanding of the movements of cells in bacteria 
or fungi colonies could be the key feature in reproducing the experimentally observed 
scaling behaviour.\\
We suggest to biologists or physicists interested in this field to undertake simple 
experiments to measure critical behaviour of colony growth  in solutions of nutrient 
at different temperatures with the intent to better understand the role of the brownian 
motion in this context and to check our results.\\
We remark the fact that this kind of evolution with 'instantaneous' chemical correlations 
and thermal movement is much more akin to the real one:  industrial or natural 
fermentation process and evolution of life very often occurs in a concentrated solution 
at room temperature rather than on an agar plate. A further step will be to implement a
new geometrical arrangement, similar to the Vicsek experiment one, with the goal to
reproduce the scaling exponents, and to study the possibility of a multilayer colony. \\
The interest for this kind of studies could be broader than what we have depicted: many 
other features arising from different chemical or biochemical mechanisms could be used 
to introduce more complexity and new patterns. 

\section*{Acknowledgement}
We would like to thank E. Orlandini for the many stimulating discussions that 
we have had on this topic. 
\\~\\
{\small email address: fortunat@pd.infn.it}

\end{document}